# The Hierarchical $\phi^4$ - Trajectory by Perturbation Theory in a Running Coupling and its Logarithm

J. Rolf[1] and C. Wieczerkowski[2]

Institut für Theoretische Physik I, Universität Münster,
Wilhelm-Klemm-Straße 9, D-48149 Münster,
[1] rolfj@planck.uni-muenster.de,
[2] wieczer@yukawa.uni-muenster.de

**Abstract**

We compute the hierarchical $\phi^4$-trajectory in terms of perturbation theory in a running coupling. In the three dimensional case we resolve a singularity due to resonance of power counting factors in terms of logarithms of the running coupling. Numerical data is presented and the limits of validity explored. We also compute moving eigenvalues and eigenvectors on the trajectory as well as their fusion rules.

# 1 Introduction

In the block spin renormalization scheme of Wilson [Wil71, KW74] renormalized theories come as renormalized trajectories of effective actions. Departing from a bare action the renormalized trajectory is reached by an infinite iteration of block spin transformations. For this limit to exist the bare couplings have to be tuned as the number of block spin transformations is increased. Consider an asymptotically free model at weak coupling. There the point is to keep couplings under control which increase by value under a block spin transformation. Such couplings are called relevant. In weakly coupled models they can be identified by naive power counting. This renormalization scheme has been beautifully implemented both within and beyond perturbation theory. Let us mention the work of Polchinski [Pol84], Gawedzki and Kupiainen [GK84], Gallavotti [Gal85], and Rivasseau [Riv91] as a guide to the extensive literature. The underlying picture of an ultraviolet asymptotically free model is to think of the renormalized trajectory as unstable manifold of a trivial fixed point. Although this picture has been in mind behind block spin renormalization since the very beginning [KW74] it has not been formalized yet to an approach free of a bare action. This paper is a contribution to fill this gap. It extends the analysis begun in [WX94] and [WX95] in the context of renormalization group improved actions for the two dimensional $O(N)$-invariant nonlinear $\sigma$-model. Here we will work it out for the $\phi^4$-trajectory in the hierarchical approximation. The hierarchical model was invented by Dyson [Dys69] and Baker [Bak72] and has enjoyed the attention of Bleher and Sinai [BS73], Collet and Eckmann [CE78], Koch and Wittwer [KW91], Felder [Fel87], and Pordt [Por93], to mention a few. The $\phi^4$-trajectory will be defined as a curve which departs the trivial fixed point in the $\phi^4$-direction. Technically we perform a renormalized perturbation expansion in a running coupling. In the three dimensional case we perform a perturbation expansion in a running coupling and its logarithm. The dynamical principle which proves to be strong enough to determine the trajectory at least in perturbation theory is stability under the renormalization group. With stability we mean that the trajectory is left invariant under a transformation as a set in theory space. Recall that a renormalized action always comes together with a sequence of descendents generated by further block spin transformations. Even in the case of a discrete transformation this sequence will prove to consist of points on a continuous curve in theory space which is stable under the renormalization group. It is the computation of this curve we address. The result is an iterative form of renormalized perturbation theory. Its closest relatives in the literature are the renormalized tree expansions of Gallavotti and collaborators [Gal85, GN85a, GN85b]. A pedagocial exposure can be found in [FHRW88]. Our expansion is however free of divergencies piled up in standard perturbation theory by infinitely iterated renormalization group transformations from the very beginning. Surprisingly we do not need to treat relevant and irrelevant couplings on a different footing. It will involve neither bare couplings nor renormalization conditions in the original sense. A renormalization group transformation in our approach translates to a transformation of the running coupling according to some $\beta$-function. We will consider in particular a choice of coordinate whose associated $\beta$-function is exactly linear. This idea has also appeared in [EW84] and references therein. Renormalized perturbation theory furthermore will surprise us with a sequence of discrete poles at special rational dimensions. These poles will be traced back to certain resonance conditions on the scaling dimensions of powers of fields. In particular the case of three dimensions will be shown to be resonant. We will resolve the associated singularity by a double expansion in both the running coupling and its logarithm. The expansion will then be extended to the computation of moving eigenvectors in the sense of [WX95] on the renormalized trajectory and their fusion rules. Finally we perform a numerical test of our renormalized actions. As expected they prove to work well in a small field region. The extension of our program to full models is under way. A prototype with momentum space regularization has been developed in [Wie].

## 2  Hierarchical renormalization group

The hierarchical renormalization group in the form advocated by Gawedzki and Kupiainen [GK84] is a theory of the non linear transformation

$$\mathcal{R}Z(\psi) = \left(\int d\mu_\gamma(\zeta)\, Z(L^{1-\frac{D}{2}}\psi + \zeta)\right)^{L^D} \tag{1}$$

on some space of Boltzmann factors $Z(\phi)$. In the scalar theory $\phi$ is a single real field variable.

$$d\mu_\gamma(\zeta) = (2\pi\gamma)^{-\frac{1}{2}} \exp\left(-\frac{\zeta^2}{2\gamma}\right) d\zeta \tag{2}$$

is the Gaussian measure on $\mathbb{R}$ with mean zero and covariance $\gamma$. The parameters of (1) are the Euclidean dimension $D$ and the block scale $L$. The subspace of even Boltzmann factors $Z(-\phi) = Z(\phi)$ is stable under (1). We will restrict our attention to this subspace. Let the potential be given by $Z(\phi) = \exp(-V(\phi))$. The transformation for the potential is

$$\mathcal{R}V(\psi) = -L^D \log\left( \int d\mu_\gamma(\zeta) \, \exp\left(-V(L^{1-\frac{D}{2}}\psi + \zeta)\right) \right) . \qquad (3)$$

The below analysis will be done in terms of the potential. The method will be perturbation theory. The question of stability bounds will not be addressed. Regarding mathematical aspects of (1) and (3) we refer to the work of Collet and Eckmann [CE78], Gawedzki and Kupiainen [GK84] and of Koch and Wittwer [KW91].

## 3 The trivial fixed point

(3) has a trivial fixed point $V_*(\phi) = 0$. This fixed point is the hierarchical massless free field. The linearization of (3) at this trivial fixed point is given by

$$\mathcal{L}_{V_*}\mathcal{R}\mathcal{O}(\psi) = L^D \int d\mu_\gamma(\zeta) \, \mathcal{O}\left(L^{1-\frac{D}{2}}\psi + \zeta\right) . \qquad (4)$$

This linearization is diagonalizable. The eigenvectors are normal ordered products

$$: \phi^n :_{\gamma\prime} = \left.\frac{\partial^n}{\partial j^n}\right|_{j=0} \exp\left( j\phi - \frac{j^2 \gamma\prime}{2} \right) \qquad (5)$$

with normal ordering covariance $\gamma\prime = \left(1 - L^{2-D}\right)^{-1} \gamma$. The normal ordering covariance has been chosen in order to be invariant with respect to integration with $d\mu_\gamma$. Its singularity at $D = 2$ is an infrared singularity of the hierarchical massless free field in two dimensions. The eigenvalues are

$$\lambda_n = L^{D+n\left(1-\frac{D}{2}\right)} . \qquad (6)$$

The eigenvalue of $: \phi^4 :_{\gamma\prime}$ is $\lambda_4 = L^{4-D}$. The eigenvector $: \phi^4 :_{\gamma\prime}$ is therefore relevant for $D < 4$, marginal for $D = 4$, and irrelevant for $D > 4$ dimensions. Perturbation theory can be used to compute corrections to (4) in a neighbourhood of $V_*(\phi)$.

# 4 The $\phi^4$ - trajectory

Let us define a curve $V(\phi, g)$ in the space of potentials parametrized by a local coordinate $g$. We call it the $\phi^4$ - trajetory. We expand the potential

$$V(\phi, g) = \sum_{n=0}^{\infty} V_{2n}(g) : \phi^{2n} :_{\gamma'} \tag{7}$$

in the base of eigenvectors (5). A natural coordinate in a vicinity of $V_*(\phi)$ is the $\phi^4$ - coupling defined by $V_4(g) = g$. Let us use it for a moment. Let the $\phi^4$ - trajectory then be the curve $V(\phi, g)$ defined by the following two conditions:

1) $V(\phi, g)$ is stable under $\mathcal{R}$. Then there exists a function $\beta(g)$ such that

$$\mathcal{R}V(\phi, g) = V(\phi, \beta(g)) . \tag{8}$$

   The function $\beta(g)$ is of course coordinate dependent. With the $\phi^4$ - coupling as coordinate it is called $\beta$ - function.

2) $V(\phi, g)$ visits the trivial fixed point $V_*(\phi)$ at $g = 0$. The tangent to $V(\phi, g)$ at $V_*(\phi)$ is given by

$$\left.\frac{\partial}{\partial g}\right|_{g=0} V(\phi, g) = : \phi^4 :_{\gamma'} \tag{9}$$

   This condition is equivalent with $V_4(g) = g + O(g^2)$ together with $V_{2n}(g) = O(g^2)$, $n \neq 2$.

The $\phi^4$ - trajectory is the object of principal interest in massless $\phi^4$ - theory at weak coupling.

# 5 Perturbation theory

The $\phi^4$ - trajectory can be computed by perturbation theory in $g$ as solution to (8) and (9). Potentials on the $\phi^4$ - trajectory are said to scale. A potential $V(\phi, g)$ is said to scale to order $s$ in g if there exists a function

$$\beta(g) = \beta^{(s)}(g) + O(g^{s+1}) ,$$
$$\beta^{(s)}(g) = \sum_{r=1}^{s} b_r g^r , \tag{10}$$

such that
$$V(\phi, g) = V^{(s)}(\phi, g) + O(g^{s+1}) ,$$
$$\mathcal{R}V^{(s)}(\phi, g) = V^{(s)}(\phi, \beta(g)) + O(g^{s+1}) , \qquad (11)$$

and
$$V^{(1)}(\phi, g) = g : \phi^4 :_{\gamma\prime} . \qquad (12)$$

The scheme is to compute $\beta^{(s+1)}(g)$ and $V^{(s+1)}(\phi, g)$ given $\beta^{(s)}(g)$ and $V^{(s)}(\phi, g)$ to some order $s$. Let us explain it in some detail at the case of $D = 4$ dimensions, block scale $L = 2$, and covariance $\gamma = 1$. Then the normal ordering covariance is $\gamma\prime = \frac{4}{3}$. Computing a block spin transformation (3), we speak of $V(\phi)$ as bare potential and of $\mathcal{R}V(\phi)$ as effective potential. The point of departure is (12). Anticipating the terms generated in $\mathcal{R}V^{(1)}(\phi, g)$ to second order in $g$ we make the ansatz

$$V^{(2)}(\phi, g) = c_0 g^2 + c_2 g^2 : \phi^2 : + g : \phi^4 : + c_6 g^2 : \phi^6 : . \qquad (13)$$

The coefficients are determined by the condition that (11) be fulfilled to second order. (13) is mapped to

$$\mathcal{R}V^{(2)}(\phi, g(g')) = (16c_0 - \frac{5440}{9})g'^2 + (4c_2 - 448)g'^2 : \phi^2 :$$
$$+ g' : \phi^4 : + (\frac{c_6}{4} - 2)g'^2 : \phi^6 : + O(g'^3) . \qquad (14)$$

Here the effective coupling defined as the coefficient of $: \phi^4 :$ in the effective potential is given by

$$g'(g) = g - 60g^2 + O(g^3) . \qquad (15)$$

Comparing the effective potential as a function of the effective coupling with the bare potential as a function of the bare coupling we conclude that

$$c_0 = \frac{1088}{27} , \; c_2 = \frac{448}{3} , \; c_6 = -\frac{8}{3} \qquad (16)$$

on the $\phi^4$ - trajectory. The coefficients of the $\beta$ - function (10) to this order are

$$b_1 = 1 , \; b_2 = -60 . \qquad (17)$$

It follows that $g$ is marginally irrelevant in four dimensions. This completes the first step. It is iterated in the obvious manner. The general form of the order $s$ approximation is

$$V^{(s)}(\phi, g) = \sum_{n=0}^{s+1} c_{2n}^{(s)}(g) : \phi^{2n} : ,$$

$$c_{2n}^{(s)}(g) = \sum_{r=2}^{s} c_{2n,r} g^r , \quad n \leqslant 1 ,$$

$$c_4^{(s)}(g) = g ,$$

$$c_{2n}^{(s)}(g) = \sum_{r=n-1}^{s} c_{2n,r} g^r , \quad n \geqslant 3 . \qquad (18)$$

It includes all normal ordered products generated in the effective potential by (3) from (12) to order $s$ in $g$. The iteration proceeds as above with the order $s + 1$ ansatz of the form (18). The condition (11) yields a system of linear equations for the order $s + 1$ coefficients. (To highest order the coefficients have no other choice.) This system has a unique solution: the $\phi^4$ - trajectory. Note that the coefficient $b_{s+1}$ of the $\beta$ - function is already determined by $V^{(s)}(\phi, g)$. For instance (15) does not contain any of the coefficients in (13). The expansion can be computed to higher orders using computer algebra. To third order we find

$$c_0^{(3)}(g) = \frac{1088}{27} g^2 - \frac{54784}{27} g^3 ,$$

$$c_2^{(3)}(g) = \frac{448}{3} g^2 - \frac{497408}{27} g^3 ,$$

$$c_6^{(3)}(g) = -\frac{8}{3} g^2 + 352 g^3 ,$$

$$c_8^{(3)}(g) = \frac{32}{3} g^3 \qquad (19)$$

together with

$$\beta^{(3)}(g) = g - 60 g^2 + 8880 g^3 . \qquad (20)$$

Let us remark that the perturbation coefficients (10) and (11) come with alternating signs. The coefficients show a frightening increase in absolute value with the order in $g$. The full series is not expected to converge. Note that the coefficients look better when $g$ is replaced by $g/4!$.

# 6 Resonances

We can apply the above scheme to compute the $\phi^4$ - trajectory in other than four dimensions. The solution is again of the form (18). We do however encounter a new phenomenon. The improvement coefficients $c_{2n}^{(s)}$ exhibit poles at certain discrete points as functions of the dimension parameter $D$. We call this phenomenon resonance because it can be traced back to the fact that certain scaling parameters become powers of one another at these points. This happens particular in three dimensions. Let us consider the transformation (3) with block scale $L = 2$ and covariance $\gamma = 1$, but this time with arbitrary value of $D$. We express the dependence on $D$ in terms of a variable $\alpha = L^D$. To third order in $g$ the improvement coefficients are given by the rational functions

$$\begin{aligned}
c_0^{(3)} &= \frac{12\alpha^3(\alpha + 4)(\alpha^2 + 16)}{(\alpha^3 - 256)(\alpha - 4)^3}g^2 \\
&\quad - \frac{288\alpha^4(\alpha^5 + 32\alpha^3 + 512\alpha^2 + 4096)(\alpha + 4)^2}{(\alpha + 8)(\alpha - 8)(\alpha^2 + 64)(\alpha^3 - 256)(\alpha - 4)^4}g^3 \,, \\
c_2^{(3)} &= \frac{48\alpha^2(\alpha^2 + 4\alpha + 16)}{(\alpha - 8)(\alpha + 8)(\alpha - 4)^2}g^2 \\
&\quad - \frac{384\alpha^3(7\alpha^5 + 46\alpha^4 + 288\alpha^3 + 1728\alpha^2 - 1024\alpha - 22528)}{(\alpha + 8)(\alpha - 8)(\alpha^3 - 1024)(\alpha - 4)^3}g^3 \,, \\
c_6^{(3)} &= -\frac{8}{3}g^2 - \frac{576\alpha(\alpha + 6)}{(\alpha - 4)(\alpha - 64)}g^3 \,, \\
c_8^{(3)} &= \frac{32}{3}g^3 \,.
\end{aligned} \qquad (21)$$

The $\beta$ - function to this order in $g$ is given by the function

$$\begin{aligned}
\beta^{(3)}(g) &= \frac{16}{\alpha}g - \frac{576(\alpha + 4)}{\alpha(\alpha - 4)}g^2 \\
&\quad + \frac{256(215\alpha^4 + 1400\alpha^3 - 10128\alpha^2)}{\alpha(\alpha - 8)(\alpha + 8)(\alpha - 4)^2}g^3 \\
&\quad - \frac{256(95744\alpha + 355328)}{\alpha(\alpha - 8)(\alpha + 8)(\alpha - 4)^2}g^3 \,.
\end{aligned} \qquad (22)$$

Singularities appear at positive dimensions

$$\begin{array}{c|ccc|cc} & \multicolumn{3}{c|}{g^2} & \multicolumn{2}{c}{g^3} \\ \hline \alpha & 4 & 256^{\frac{1}{3}} & 8 & 1024^{\frac{1}{3}} & 64 \\ \hline D & 2 & \frac{8}{3} & 3 & \frac{10}{3} & 6. \end{array} \qquad (23)$$

As one goes to higher orders in $g$ more and more poles show up in (21) and (22). Let us have a closer look at the three dimensional pole to second order in $g$. Inserting (13) into (3) at three dimensions we find

$$\begin{aligned} \mathcal{R} V^{(2)}(\phi, g(g')) &= (2c_0 - 360) + (c_2 - 336)\, g'^2 : \phi^2 : \\ &\quad + g' : \phi^4 : + \left(\frac{c_6}{4} - 2\right) g'^2 : \phi^6 : + O(g'^3)\ . \end{aligned} \qquad (24)$$

The parameters of (3) are $D = 3$, $L = 2$, and $\gamma = 1$. The normal ordering covariance is $\gamma\prime = 2\gamma$. The $\beta$ - function to this order is given by

$$g'(g) = 2g - 216 g^2 + O(g^3)\ . \qquad (25)$$

The $:\phi^4:$ - coupling is therefore relevant in three dimensions. From (24) we would conclude that

$$c_0 = 360\ ,\ c_6 = -\frac{8}{3}\ . \qquad (26)$$

But there exists no solution to the equation for $c_2$ (besides infinity). It follows that the $:\phi^2:$ - coupling cannot be written in terms of a power series in the $:\phi^4:$ - coupling on the $\phi^4$ - trajectory in three dimensions. The point is that the $:\phi^2:$ - coupling flows like

$$-336 n 2^{2n} = -\frac{336}{\log(2)} \log(g) g^2 \qquad (27)$$

with $g = 2^n$ upon iteration of (24). This suggests a double expansion in both $g$ and $\log(g)$.

# 7   Linear $\beta$ - function

So far we have used the $\phi^4$ - coupling as coordinate for the $\phi^4$ - trajectory. It was defined by the condition $V_4(g) = g$ following (7). This coordinate leads to unnecessary complications when dealing with its logarithm. A better coordinate which is also interesting by itself is the linear coordinate defined by the condition that the $\beta$ - function be exactly given by

$$\beta(g) = L^{4-D}g \qquad (28)$$

on the $\phi^4$ - trajectory. In terms of this linear $\beta$ - function the transformation on the $\phi^4$ - trajectory looks exactly like the linearized renormalization group. The definitions (8) and (9) remain untouched by (28). Using the expansion with the linear $\beta$ - function

$$V_4(g) = g + \sum_{r=2}^{\infty} V_{4,r} g^r \qquad (29)$$

also becomes a computable power series. The strategy is the same as above. For the case of $L = 2$ and $\gamma = 1$ the $\phi^4$ - trajectory is given by

$$\begin{aligned}
V_0^{(3)} &= \frac{12\alpha^3(\alpha+4)(\alpha^2+16)}{(\alpha^3-256)(\alpha-4)^3}g^2 \\
&\quad + \frac{576\alpha^4(\alpha^3+8\alpha^2+8\alpha+256)(\alpha+4)^2}{(\alpha-16)(\alpha+8)(\alpha-8)(\alpha^2+64)(\alpha-4)^4}g^3 \;, \\
V_2^{(3)} &= \frac{48\alpha^2(\alpha^2+4\alpha+16)}{(\alpha-8)(\alpha+8)(\alpha-4)^2}g^2 \\
&\quad + \frac{768\alpha^3(\alpha^6+69\alpha^5+368\alpha^4-2880\alpha^3)}{(\alpha-16)(\alpha+8)(\alpha-8)(\alpha^3-1024)(\alpha-4)^3}g^3 \\
&\quad + \frac{768\alpha^3(-22528\alpha^2-144384\alpha-475136)}{(\alpha-16)(\alpha+8)(\alpha-8)(\alpha^3-1024)(\alpha-4)^3}g^3 \;, \\
V_4^{(3)} &= g + \frac{36\alpha(\alpha+4)}{(\alpha-4)(\alpha-16)}g^2 \\
&\quad - \frac{16\alpha^2(53\alpha^5-3336\alpha^4-24752\alpha^3)}{(\alpha+8)(\alpha-8)(\alpha+16)(\alpha-16)^2(\alpha-4)^2}g^3 \\
&\quad - \frac{16\alpha^2(149248\alpha^2+1342464\alpha+5685248)}{(\alpha+8)(\alpha-8)(\alpha+16)(\alpha-16)^2(\alpha-4)^2}g^3 \;, \\
V_6^{(3)} &= -\frac{8}{3}g^2 - \frac{384\alpha(2\alpha^2-45\alpha-272)}{(\alpha-4)(\alpha-16)(\alpha-64)}g^3 \;, \\
V_8^{(3)} &= \frac{32}{3}g^3 \qquad (30)
\end{aligned}$$

To each order of perturbation theory we find a system of linear equations which has a unique solution. The coefficients are again rational functions in $\alpha = L^D$ with poles at resonant dimensions. Note that (30) has an additional pole at $D = 4$ as compared with (21). In four dimensions (28) becomes the identity and (8) a fixed point equation. Resonances are now easily understood. To order $m$ in $g$ the transformation (3) acts as

$$cg^m : \phi^{2n} : \mapsto L^{D+n(2-D)} cg^m : \phi^{2n} :$$
$$= L^{D+n(2-D)-m(4-D)} c\beta(g)^m : \phi^{2n} : . \qquad (31)$$

A resonance thus occurs if

$$D + n(2-D) - m(4-D) = 0 . \qquad (32)$$

In the case of $D = 3$ dimensions this condition becomes

$$3 - n - m = 0 . \qquad (33)$$

Since $m \geqslant 2$ in this business we find only two resonant terms $(n, m)$ in three dimensions: $(1, 2)$ and $(0, 3)$. The former is a mass resonance, the latter a vacuum resonance. The resonant dimensions are rational and given by

$$D = \frac{4m - 2n}{1 - n + m} . \qquad (34)$$

In particular table (23) is immediatly reproduced. An interesting variation of the linear $\beta$ - function consists of replacing (28) by

$$\begin{aligned} \beta(g) &= b_1 g + b_2 g^2 , \\ b_1 &= L^{4-D} , \\ b_2 &= 36 L^{4-D} \frac{L^2 + L^D}{L^2 - L^D} , \end{aligned} \qquad (35)$$

truncating the $\beta$ -function (8) to second (or more generally to any fixed) order of perturbation theory in the $\phi^4$ - coupling $g$. This $\beta$ - function has a fixed point at finite $g$ for $D < 4$. Supposing that our expansion was summable in some sense it follows that the nontrivial fixed point of Koch and Wittwer is on the $\phi^4$ - trajectory. Recall that in these coordinates (8) holds exactly with $\beta$ - function (35). It follows that a fixed point of (35) is a fixed point of (3). We will not pursue this line of thought further at this instant. The $\beta$ - function (28) is ideally suited for the double expansion in both $g$ and $\log(g)$.

# 8 Perturbation theory in $g$ and $\log(g)$

Let us consider the $\phi^4$ - trajectory in $D = 3$ dimensions, again with $L = 2$ and $\gamma = 1$. The problem that $c_2$ cannot be determined in (24) such that it remains invariant to second order can be cured as follows. We use the linear $\beta$ - function defined by (28) and expand the potential in both $g$ and

$$\kappa = \log(g) \ . \tag{36}$$

In the expansion we treat $\kappa$ as an independent variable having the same order as $g^0$. Note however that our expansion consists of at least differentiable combinations of $g$ and $\kappa$. To second order in $g^2$ we can replace (13) by the ansatz

$$V^{(2)}(\phi, g) = c_0 g^2 + (c_2 + c_{2,1}\kappa)g^2 : \phi^2 : + (g + c_4 g^2) : \phi^4 : + c_6 g^2 : \phi^6 : \ . \tag{37}$$

Here we have left out all terms which anyway turn out to be zero on the $\phi^4$ - trajectory. In terms of $g' = 2g$ and $\kappa' = \kappa + \log(2)$ the ansatz (37) is mapped by (3) to

$$\begin{aligned}
\mathcal{R}V^{(2)}(\phi, g(g')) &= (2c_0 - 360) \, g'^2 \\
&+ (c_2 - \log(2)c_{2,1} - 336 + c_{2,1}\kappa') \, g'^2 : \phi^2 : \\
&+ \left(g' + \left(\frac{c_4}{2} - 54\right) g'^2\right) : \phi^4 : \\
&+ \left(\frac{c_6}{4} - 2\right) g'^2 : \phi^6 : + O(g'^3) \ . 
\end{aligned} \tag{38}$$

As a consequence (37) reproduces its form up to a change of the running coupling (28) if and only if

$$c_0 = 360 \ , \ c_4 = -108 \ , \ c_6 = -\frac{8}{3} \ , \ c_{2,1} = -\frac{336}{\log(2)} \ . \tag{39}$$

The parameter $c_2$ is free. To second order in $g$ we find a one parameter family of solutions to (8) and (9). The free parameter is associated with the mass resonance $(1, 2)$ of (33). One immediately anticipates another free parameter to third order in $g$ coming with the vacuum resonance $(0, 3)$. This is indeed the case. The general solution of (8) and (9) to third order in $g$ is given by

$$\begin{aligned}
V^{(3)}(\phi, g) &= 360g^2 + \frac{54432}{\log(2)}g^3\kappa + c_0 g^3 \\
&+ \left(c_2 g^2 - \frac{336}{\log(2)}g^2\kappa + (116928 - 36c_2)g^3 + \frac{12096}{\log(2)}g^3\kappa\right) :\phi^2: \\
&+ \left(g - 108g^2 + (17520 - \frac{8}{3}c_2)g^3 + \frac{896}{\log(2)}g^3\kappa\right) :\phi^4: \\
&+ \left(-\frac{8}{3}g^2 + 864g^3\right) :\phi^6: \\
&+ \frac{32}{3}g^3 :\phi^8: \qquad (40)
\end{aligned}$$

including $c_0$ and $c_2$ as free parameters. Thereafter we have no further free parameters in the higher order coefficients. Thus we have to supplement the definition of the $\phi^4$ - trajectory given by (8) and (9) by two additional conditions on $c_0$ and $c_2$ in three dimensions to single out a curve in the space of interactions. We choose

$$c_0 = 0 \, , \, c_2 = 0 \, . \qquad (41)$$

With this choice the perturbation theory has a minimal number of vertices. (41) can be thought of as additional renormalization conditions.

At higher orders the scheme explained above iterates. The general form of the potential at order $s$ is

$$\begin{aligned}
V^{(s)}(\phi, g) &= \sum_{n=0}^{s+1} V_{2n}^{(s)}(g) :\phi^{2n}: \, , \\
V_{2n}^{(s)}(g) &= \sum_{r=2}^{s}\sum_{t=0}^{[r/2]} V_{2n,r,t} g^r \kappa^t \, , \quad n \leqslant 1 \, , \\
V_4^{(s)}(g) &= g + \sum_{r=2}^{s}\sum_{t=0}^{[r/2]} V_{4,r,t} g^r \kappa^t \, , \\
V_{2n}^{(s)}(g) &= \sum_{r=n-1}^{s}\sum_{t=0}^{[r/2]} V_{2n,r,t} g^r \kappa^t \, , \quad n \geqslant 3 \, . \qquad (42)
\end{aligned}$$

Here the third order coefficients are given by (40) and, for instance, (41). To each further order of perturbation theory we meet a system of linear equations possessing a unique solution for the coefficients in (42): the $\phi^4$ - trajectory in terms of the double expansion. Recall that $\kappa$ should be substituted by $\log(g)$ in (42).

# 9 Numerical calculation of the renormalized trajectory

Hierarchical renormalization group flows can also be computed using standard numerical methods. Let us perform a numerical investigation of the transformation (3) in order to determine the limits of validity of the expansion (42) in $g$ and $\log(g)$. Furthermore we want to investigate the large field behaviour of our expansion. We choose the following numerical setup. To calculate the transformation (3) iteratively we sample the potential $V$ at $N$ equidistant points betwen 0 and $\phi_{max}$. Then we perform a cubic spline interpolation and integrate using standard NAG library functions. To reduce the error due to boundary effects at $\phi = \phi_{max}$ we always choose $\phi_{max}$ so that $V(\phi_{max}) = V_{max}$ with $V_{max}$ large enough; for example $V_{max} = 20$ will do. For $\phi > \phi_{max}$ we set

$$V(\phi) = V_{HT}(\phi - a) - b \qquad (43)$$

and choose $a$ and $b$ so that the first derivative of $V$ at $\phi_{max}$ is continuous. Here with $V_{HT}$ we mean the quadratic high temperature fixed point of (3). Field asymptotics have been investigated by Koch and Wittwer in their work on the nontrivial double well fixed point [KW91]. That means that we supplement our numerical calculation with the expectation that the potentials on the renormalized trajectory have HT-like asymptotic behaviour. We let the fluctuation field $\zeta$ vary between $-\zeta_{max}$ and $\zeta_{max}$ and choose $\zeta_{max} = 20$. This produces errors which can be neglected. As in the case of the expansion we restrict our attention to the room of symmetric potentials which is invariant under (3). All potentials were calculated with $L = 2$, $D = 3$, $\gamma = 1$ and $N = 401$. To determine the renormalized trajectory according to our definition in section 4 supplemented by the condition $c_2 = 0$, see section 8, we proceed as follows. We start with a bare potential

$$V = \left(\frac{336}{\log(2)}g_0{}^2\kappa + 116928 g_0{}^3 + \frac{12096}{\log(2)}g_0{}^3\kappa\right) :\phi^2:$$
$$+ \left(g_0 - 108 g_0{}^2 + 17520 g_0{}^3 + \frac{896}{\log(2)}g_0{}^3\kappa\right) :\phi^4:$$
$$+ \left(-\frac{8}{3}g_0{}^2 + 864 g_0{}^3\right) :\phi^6:$$
$$+ \frac{32}{3}g_0{}^3 :\phi^8: \tag{44}$$

and choose $g_0$ to be a sufficiently small number, that means that an iteration starting with $L^{-m}g_0$ would yield the same trajectory if $m \in \mathbb{N}$. We have chosen $g_0 = 10^{-6}$. Note that we have to choose the third order of our perturbation expansion as input because the second order wouldn't be stable under iterated renormalization group transformations because of the large field behaviour. It is convenient to normalize the potential with the condition $V(\phi = 0) = 0$. The numerically determined potential will be refered to as exact in the following. We perform 12 iterations of the transformation (3). The analogous perturbative potentials can be found by solving the equation

$$g_0 = V_4(\tilde{g}_0) \tag{45}$$

for $\tilde{g}_0$ (which gives approximately $10^{-6}$ of course) and then follow the renormalization group flow to

$$L^n \tilde{g}_0, \quad n = 0, \ldots, 13 . \tag{46}$$

For the calculation of the perturbative potentials we used the seventh order of our perturbation expansion. The comparison between the exact and the perturbative potentials can be seen in the following pictures. The exact potentials correspond to the continuous lines the perturbative potentials correspond to the dashed lines. In figure (1) one can see that after 8 renormalization group steps the exact and the perturbative data are nearly the same. If however the number of renormalization group steps exceeds 9, there is a clear digression for $\phi \geqslant 18$. This can be seen in figure (2). That means that the large field behaviour of the perturbative expansion is wrong, since the asymptotic behaviour of the exact potential is quadratic. If the number of renormalization group steps is bigger than 10 the perturbative expansion is only valid for very small fields. We therefore conclude that our expansion is not valid any more for $g \gtrsim 10^{-3}$. The borderline has been illustrated in figure (3). One can however improve the perturbative data if one is willed to work with padé approximants. In figure (4) we compare the $(9,7)$ padé approximant of our perturbation series in the variable $\phi$ ( which has the expected asymptotic behaviour ) with the numerical data after 11 renormalization group steps. Amazingly enough both curves coincide. We will not go in details here. Let us only mention that some padé approximants develop additional unphysical poles at real values of $\phi$.

## 10  Observables

Consider a local observable $\mathcal{O}(\phi(x))$ in a full theory . In the hierarchical approximation it corresponds to an observable function $\mathcal{O}(\phi), \phi \in \mathbb{R}$. Observables are transformed according to the linearized renormalization group. The linearization $\mathcal{L}_V \mathcal{R} \mathcal{O}$ of the transformation (3) in the direction $\mathcal{O}$ at the potential $V$, defined by

$$\mathcal{L}_V \mathcal{R} \mathcal{O}(\psi) = \left.\frac{\partial}{\partial z}\right|_{z=0} \mathcal{R}(V + z\mathcal{O})(\psi) \qquad (47)$$

is given by

$$\mathcal{L}_V \mathcal{R} \mathcal{O}(\psi) = \frac{\int d\mu_\gamma(\zeta) \; \mathcal{O}(L^{1-\frac{D}{2}}\psi + \zeta) \exp\left(-V(L^{1-\frac{D}{2}}\psi + \zeta)\right)}{\int d\mu_\gamma(\zeta) \; \exp\left(-V(L^{1-\frac{D}{2}}\psi + \zeta)\right)} \qquad (48)$$

after division by $L^D$. Note that this is analogous to the linear block spin transformation of local observables in the full model. Now we want to find the running eigenvectors and eigenvalues of $\mathcal{L}_V \mathcal{R} \mathcal{O}$ along the renormalized trajectory. Here an observable is called a running eigenvector if it satisfies the equation

$$\mathcal{L}_{V_{RT}} \mathcal{R} \mathcal{O}(\phi, g) = e(\beta(g))\mathcal{O}(\phi, \beta(g)) . \qquad (49)$$

$e(g)$ is the corresponding running eigenvalue. Note that $\mathcal{O}$ and $e$ are parametrized by the perturbative expansion parameter $g$. $\beta(g)$ means the $\beta$ - function, for example (28). We already know the perturbative expansion of the potential $V$ along the renormalized trajectory (in a certain neighbourhood of the trivial fixed point) so all we have to do is to solve (49) for $\mathcal{O}$ perturbatively order by order. We start our iteration with the eigenvectors $: \phi^n :_{\gamma'}$ at the trivial fixed point, see section 3. The improved observable to the order $s$ of perturbation theory which equals $: \phi^n :_{\gamma'}$ for $g = 0$ will be called $\mathcal{O}_n^{(s)}(\phi, g)$. The corresponding running eigenvalue will be denoted by $e_n^{(s)}(g)$. Let us explain the scheme of the iteration to some detail by calculating $\mathcal{O}_2^{(1)}(\phi, g)$ in three dimensions, using the linear $\beta$-function. To first order of perturbation theory for this observable we expect

$$\mathcal{O}_2^{(1)}(\phi, g) = c_{4,0} g : \phi^4 : + : \phi^2 : + (c_{0,0} + c_{0,1}\kappa) g . \tag{50}$$

We now calculate an effective observable following the transformation (48) to get

$$\begin{aligned}
\mathcal{L}_{V_{RT}} \mathcal{R} \mathcal{O}_2^{(1)}(\phi, g) &= \left(\frac{1}{8} c_{4,0} - 1\right) g : \phi^4 : \\
&+ \left(\frac{1}{2} - 9g\right) : \phi^2 : \\
&+ \left(\frac{1}{2} c_{0,0} + \frac{1}{2} c_{0,1} \left(\kappa - \ln(2)\right)\right) g .
\end{aligned} \tag{51}$$

Therefore we conclude that the running eigenvalue $e_2^{(1)}(g)$ to this order is given by

$$e_2^{(1)}(g) = \frac{1}{2} - 9g . \tag{52}$$

Now we divide by $e_2^{(1)}(g)$ and get

$$\begin{aligned}
\mathcal{O}_{2,eff}^{(1)} &= \left(\frac{1}{4} c_{4,0} - 2\right) g : \phi^4 : \\
&+ : \phi^2 : \\
&+ (c_{0,0} + c_{0,1} \left(\kappa - \ln(2)\right)) g ,
\end{aligned} \tag{53}$$

after scaling of $g$. That means that we have the desired invariance if we set

$$c_{4,0} = -\frac{8}{3} , \quad c_{0,1} = 0 . \tag{54}$$

Analogous to the determination of the potential along the renormalized trajectory we find that one parameter, $c_{0,0}$, can be freely chosen. This can be viewed as an extra renormalization condition which is imposed on the observable $\mathcal{O}_2$. In the general case this occurs under the following circumstances: Let us have a look at a term $cg^l : \psi^m :$ of the observable $\mathcal{O}_n$ to $l$-th order of perturbation theory. This is mapped to

$$cL^{m\left(1-\frac{D}{2}\right)+l(D-4)-n\left(1-\frac{D}{2}\right)}g^l : \psi^m : \tag{55}$$

because we divide by $e_n(g)$. That means that resonances occur if

$$D = \frac{8l + 2n - 2m}{2l - m + n} \tag{56}$$

or

$$m = \frac{2l(4-D)}{2-D} + n. \tag{57}$$

For the case $D = 3$ this means that we have extra renormalization conditions whenever

$$m = n - 2l . \tag{58}$$

So for the observable $\mathcal{O}_4$ for example we have two such conditions.
In the following table we display three observables using the parameters $L = 2$, $D = 3$, $\gamma = 1$. Here and in the rest of the article we choose all free parameters to be zero.

$$\begin{aligned}
\mathcal{O}_0^{(3)}(\phi,g) &= 1, \\
\mathcal{O}_1^{(3)}(\phi,g) &= :\phi:_{\gamma\prime} \\
&\quad + \left(-\frac{4}{3}g + 144g^2 - 896\frac{\kappa g^3}{\ln(2)} - 23360 g^3\right):\phi^3:_{\gamma\prime} \\
&\quad + \left(\frac{16\,g^2}{3} - 1728\,g^3\right):\phi^5:_{\gamma\prime} \\
&\quad - \frac{256\,g^3}{9}:\phi^7:_{\gamma\prime}, \\
\mathcal{O}_2^{(3)}(\phi,g) &= -960\,g^2 + \frac{2016\,\kappa g^2}{\ln(2)} - 632448\,g^3 \\
&\quad + :\phi^2:_{\gamma\prime} \\
&\quad + \left(-\frac{8\,g}{3} + 432\,g^2 - \frac{1792\,g^3\kappa}{\ln(2)} - \frac{483072\,g^3}{5}\right):\phi^4:_{\gamma\prime} \\
&\quad + \left(\frac{112\,g^2}{9} - \frac{164160\,g^3}{31}\right):\phi^6:_{\gamma\prime} \\
&\quad - \frac{640\,g^3}{9}:\phi^8:_{\gamma\prime}.
\end{aligned} \qquad (59)$$

The corresponding eigenvalues are

$$\begin{aligned}
e_0^{(3)} &= 1, \\
e_1^{(3)} &= \frac{\sqrt{2}}{2} + \frac{84\,\sqrt{2}\kappa g^2}{\ln(2)} + 28\,\sqrt{2}g^2 - \frac{3528\,\sqrt{2}\kappa\,g^3}{\ln(2)} - 34608\,\sqrt{2}g^3, \\
e_2^{(3)} &= 1/2 - 9\,g + \frac{168\,\kappa g^2}{\ln(2)} + 1232\,g^2 - \frac{17136\,g^3\kappa}{\ln(2)} - 287736\,g^3. \qquad (60)
\end{aligned}$$

We have calculated the running eigenvectors and the running eigenvalues to seventh order. This is used for the fusion rules which will be derived in the next section and for the comparison with the numerical data.

## 11 Fusion rules

With the help of the observables $\mathcal{O}_n$ on the renormalized trajectory one can compute more general correlation functions. This task requires the introduction of a new concept, the hierarchical fusion rules. Given two observables

$\mathcal{O}_n(\phi, g)$ and $\mathcal{O}_m(\phi, g)$ these are defined by

$$\mathcal{O}_n(\phi, g)\mathcal{O}_m(\phi, g) = \sum_{l=0}^{\infty} N_{nm}^l(g)\mathcal{O}_l(\phi, g) . \qquad (61)$$

Note that with the coefficient $N_{nm}^0$ we get a symmetric bilinear form $<,>$ on the space of observables, defined by

$$< \mathcal{O}_n(\phi, g), \mathcal{O}_m(\phi, g) > := N_{nm}^0(g) . \qquad (62)$$

In the thermodynamic limit only the overlap of an observable with the constant term will survive.
According to our general concept we expand the fusion coefficients $N_{nm}^l(g)$ perturbatively:

$$N_{nm}^{l\,(s)}(g) = \sum_{k=0}^{s} N_{nm\,(k)}^l g^k . \qquad (63)$$

Now it is straightforward to calculate the fusion coefficients to some order of perturbation theory. To zeroth order we recover the well known fusion rules for normal ordered products ,of course, for example

$$\begin{aligned} \mathcal{O}_1^{(0)}(\phi, g)\mathcal{O}_1^{(0)}(\phi, g) &= 2\mathcal{O}_0^{(0)}(\phi, g) + \mathcal{O}_2^{(0)}(\phi, g) , \\ \mathcal{O}_2^{(0)}(\phi, g)\mathcal{O}_2^{(0)}(\phi, g) &= 8\mathcal{O}_0^{(0)}(\phi, g) + 8\mathcal{O}_2^{(0)}(\phi, g) + \mathcal{O}_4^{(0)}(\phi, g) . \end{aligned} \qquad (64)$$

Here the parameters are $D = 3$, $\gamma = 1$, and therefore the normal ordering covariance $\gamma\prime$ equals 2. To second order we get for example

$$\begin{aligned}
\mathcal{O}_1^{(2)}(\phi,g)\mathcal{O}_1^{(2)}(\phi,g) &= \left(2 + g^2\left(-2016\frac{\kappa}{\ln(2)} + \frac{3136}{3}\right)\right)\mathcal{O}_0^{(2)}(\phi,g) \\
&\quad + \left(1 - 16g + 1856g^2\right)\mathcal{O}_2^{(2)}(\phi,g) \\
&\quad - 48g^2\mathcal{O}_4^{(2)}(\phi,g) \,, \\
\mathcal{O}_1^{(2)}(\phi,g)\mathcal{O}_2^{(2)}(\phi,g) &= \left(4 + g\left(168\frac{\kappa}{\ln(2)} - 32\right)\right. \\
&\quad \left. + g^2\left(8736\frac{\kappa}{\ln(2)} - \frac{165856}{3}\right)\right)\mathcal{O}_1^{(2)}(\phi,g) \\
&\quad + \left(1 - 32g + g^2\left(224\frac{\kappa}{\ln(2)} + 5504\right)\right)\mathcal{O}_3^{(2)}(\phi,g) \\
&\quad - 96g^2\mathcal{O}_5^{(2)}(\phi,g) \,, \\
\mathcal{O}_2^{(2)}(\phi,g)\mathcal{O}_2^{(2)}(\phi,g) &= \left(8 - 720g\right. \\
&\quad \left. + g^2\left(-179424\frac{\kappa}{\ln(2)} + \frac{169472}{3}\right)\right)\mathcal{O}_0^{(2)}(\phi,g) \\
&\quad + \left(8 + g\left(672\frac{\kappa}{\ln(2)} - 256\right)\right. \\
&\quad \left. + 8g^2\left(69888\frac{\kappa}{\ln(2)} - \frac{917312}{3}\right)\right)\mathcal{O}_2^{(2)}(\phi,g) \\
&\quad + \left(1 - 64g + g^2\left(1792\frac{\kappa}{\ln(2)} + 14720\right)\right)\mathcal{O}_4^{(2)}(\phi,g) \\
&\quad - 192g^2\mathcal{O}_6^{(2)}(\phi,g) \,. \qquad (65)
\end{aligned}$$

Note that we have chosen the free parameters to be zero in these formulas.

## 12 Numerical calculation of the observables and eigenvalues

Similar to the potentials on the renormalized trajectory we are now going to calculate the eigenvalues and observables of our theory in three dimensions numerically. The aim is again the determination of the region of validity of our peturbative expansion in $g$ and $\log(g)$ now for the eigenvalues and observables. To this end we restrict the action of the linearized renormalization group transformation (48) to the finite dimensional space of observables which is spanned by

$$\phi^m , \quad 0 \leqslant m \leqslant M .\tag{66}$$

After performing an expansion (which is most conveniently done by differentiation) we get a finite dimensional representation matrix $L$ of the linearized renormalization group transformation

$$\mathcal{L}_{V_{RT}}\mathcal{R}\phi^i = \sum_{j=0}^{M} L_{i,j}\phi^j .\tag{67}$$

Now we are able to calculate the eigenvalues and observables of $L$. Because the matrix $L$ is of course only an approximation of the transformation (48) we have to choose $M$ big enough to get correct results. We have chosen $M = 8$ and expect that the first four eigenvalues and eigenvectors do not suffer from big errors due to the trucation. In figure (5) we have plotted the four largest eigenvalues of transformation (48) against the number of renormalization group steps. The crosses correspond to the exact (i.e. numerical) calculation whereas the boxes correspond to our perturbation expansion. We start at the perturbative potential at $g_0 = 10^{-6}$, that means near the trivial fixed point. Then we follow the renormalized trajectory performing numerical and perturbative renormalization group steps. After six renormalization group steps, i.e. at $g = 2^6 g_0$ perturbatively, compare section 7, one can see the first small deviations between the perturbative and the exact eigenvalues. After 10 renormalization group steps (i.e. at $g = 1.0 \cdot 10^{-3}$) there is a clear distinction between both. We recover and sharpen the former result that for $g \gtrsim 10^{-3}$ our expansion cannot be said to be valid any more because of nonperturbative effects.

In figure (6) to figure (9) we have plotted the exact and perturbative observables $\mathcal{O}_2$ and $\mathcal{O}_3$ after four and after six renormalization group steps, respectively. Here the perturbative data correspond to the dashed lines. For large fields we have deviations for both observables which are in some extent due to the truncation of the transformation (48). The small field behaviour has been replotted in the small pictures respectively in order to illustrate the more significant influence of the truncation on the observable $\mathcal{O}_3$ as well as to illustrate the right behaviour at $\phi = 0$ of our calculations. Here we just recover the approximate normal ordering near the trivial fixed point. Because of nonperturbative effects we cannot reach the nontrivial fixed point using the naive series for the eigenvalues. It would be very interesting however to find a way to sum up our expansion in $g$ and $\log(g)$ in order to calculate critical exponents at the nontrivial fixed point. This problem will be investigated in our future work.

# Acknowledgement

We are grateful to Klaus Pinn , Andreas Pordt and York Xylander for helpful discussions. J.R. would like to thank the Studienstiftung des deutschen Volkes for financial support.

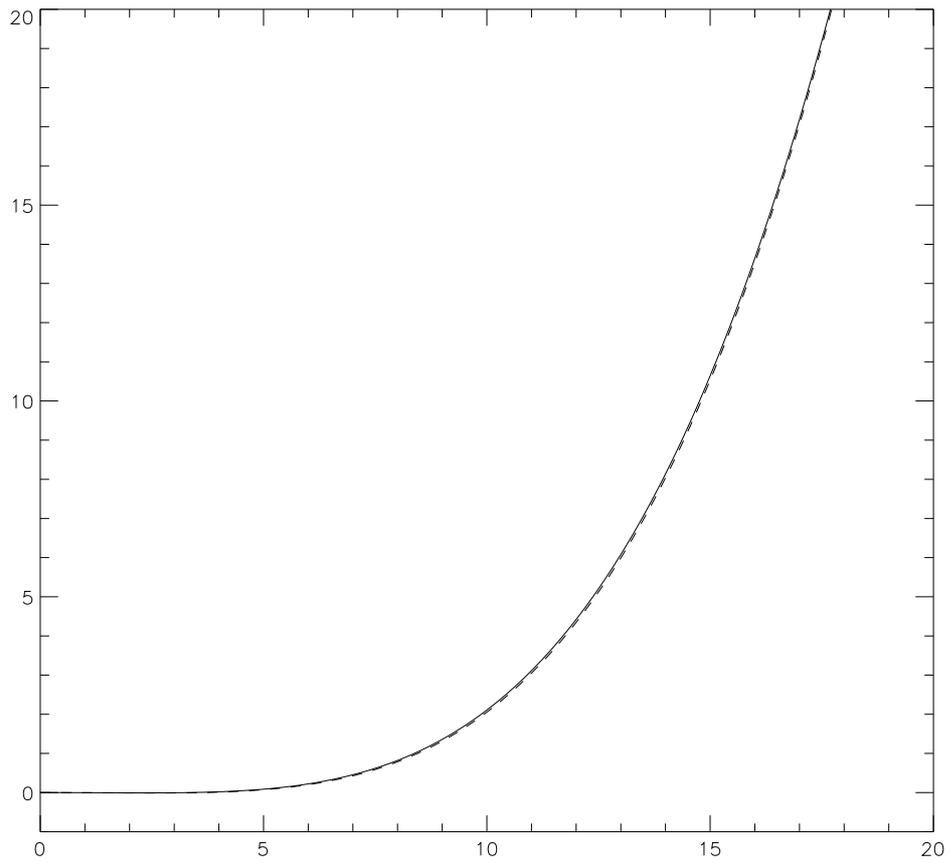

Figure 1: exact and perturbative potential after 8 renormalization group steps

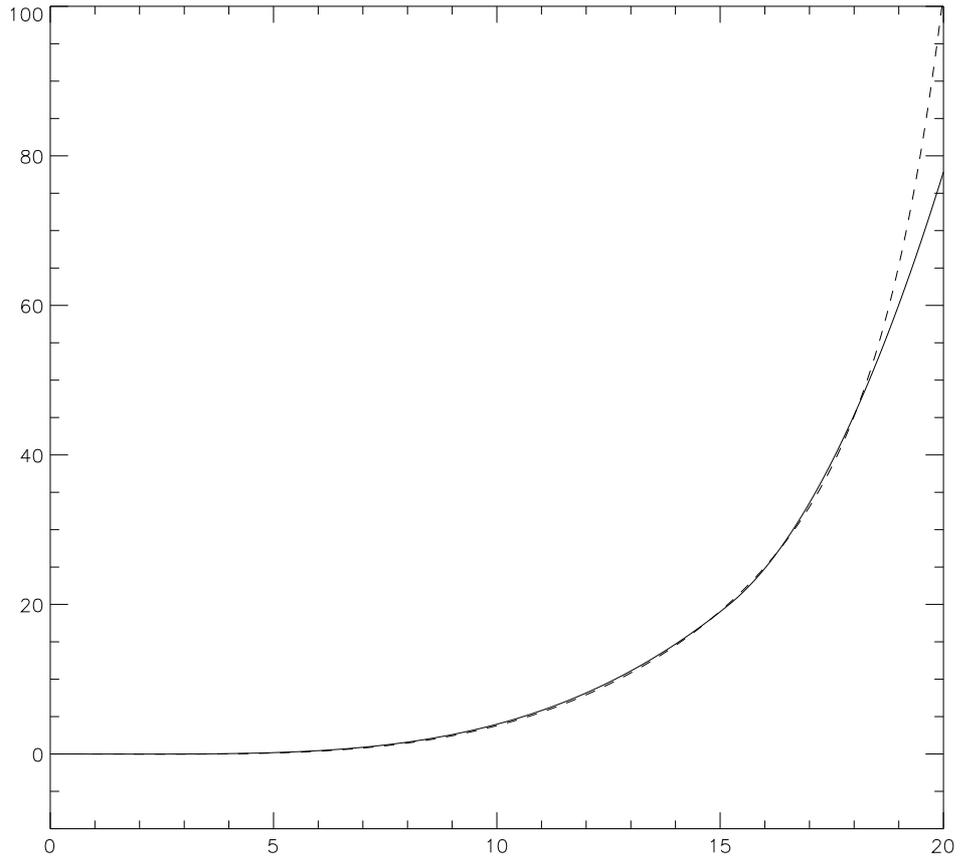

Figure 2: exact and perturbative potential after 9 renormalization group steps

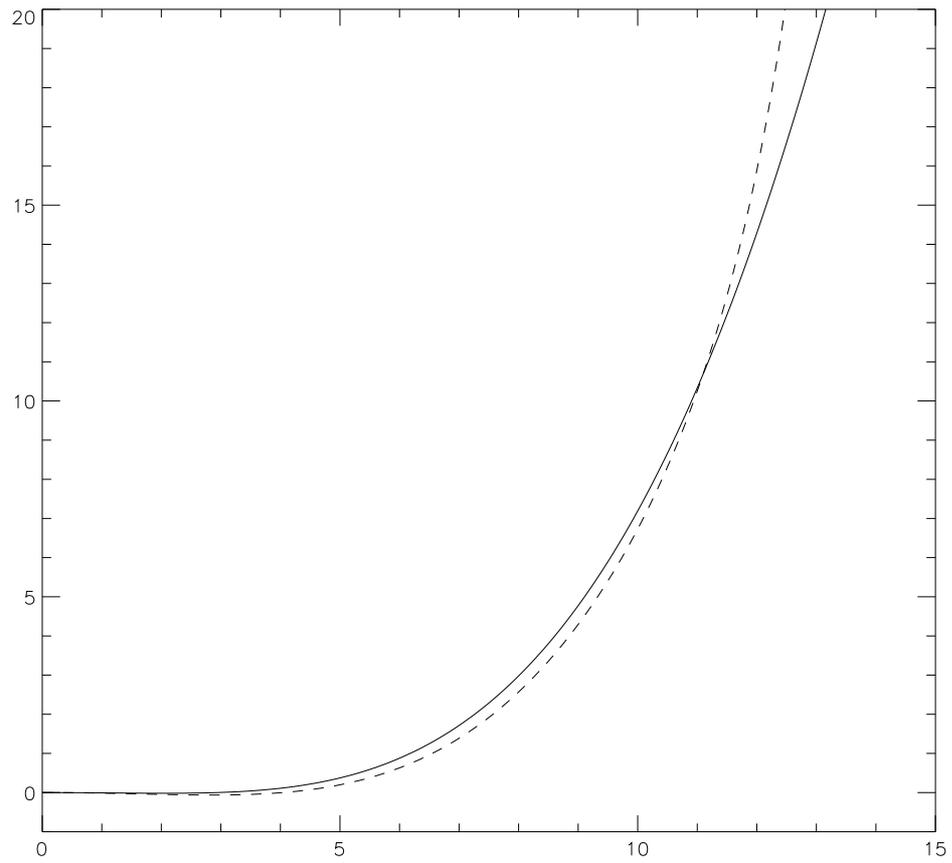

Figure 3: exact and perturbative potential after 10 renormalization group steps

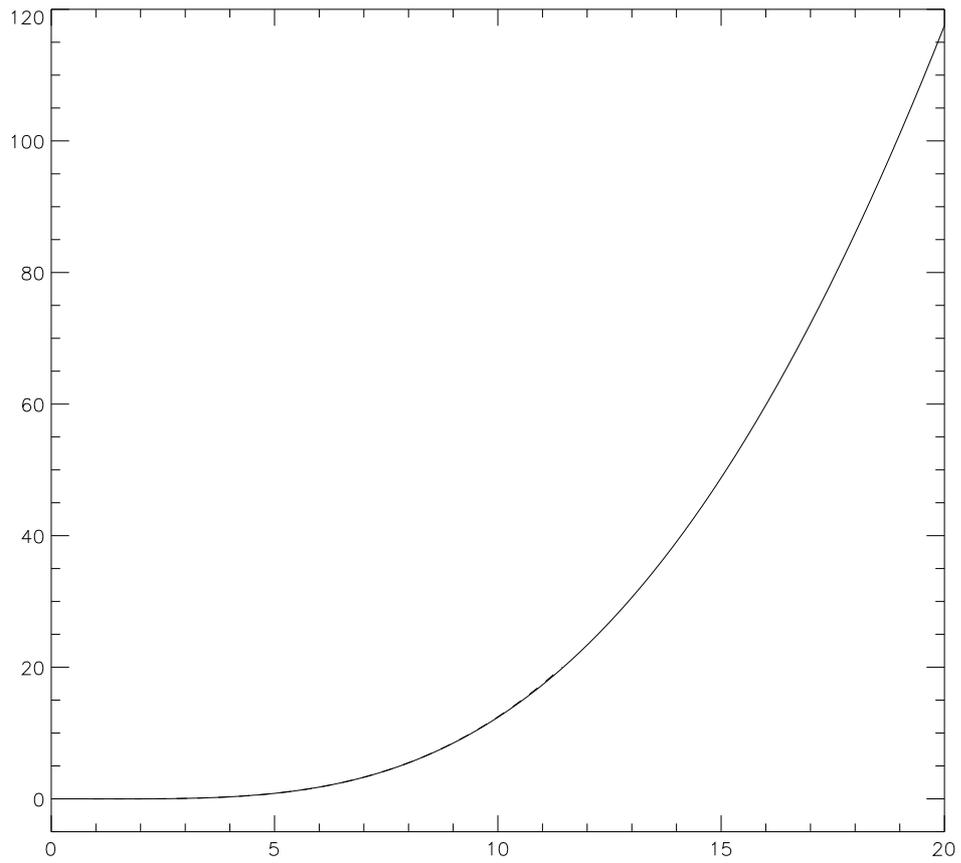

Figure 4: exact potential and padé approximant of the perturbative potential after 11 renormalization group steps

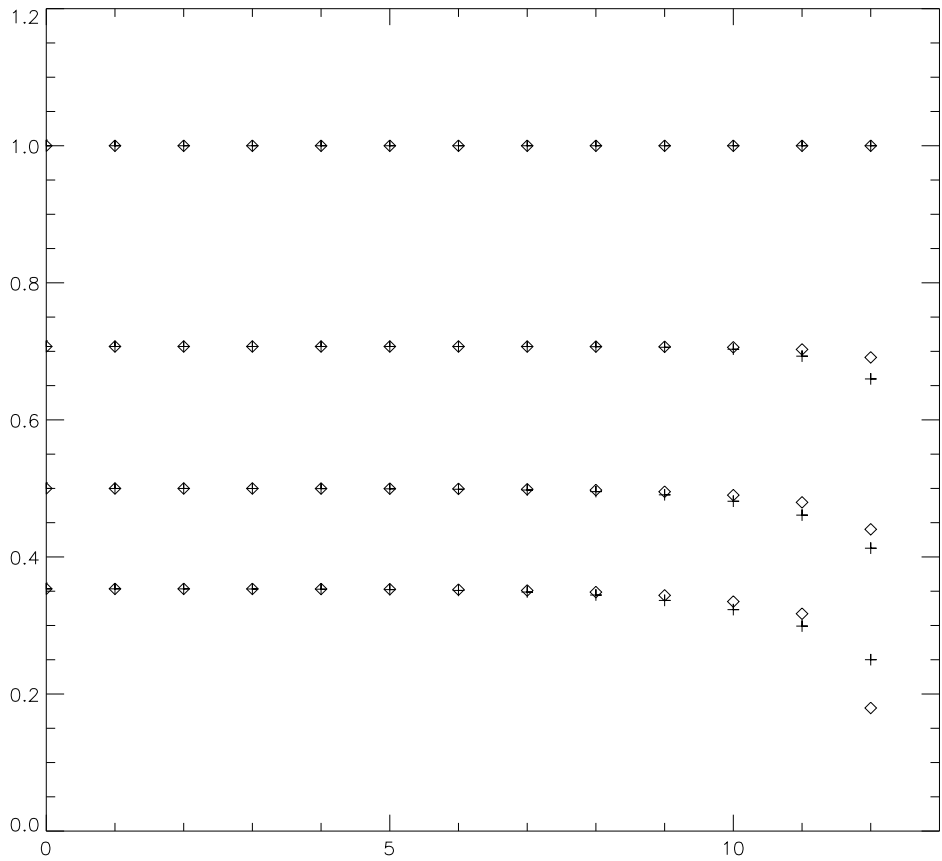

Figure 5: exact (crosses) and perturbative (boxes) eigenvalues of transformation (48) against number of renormalization group steps

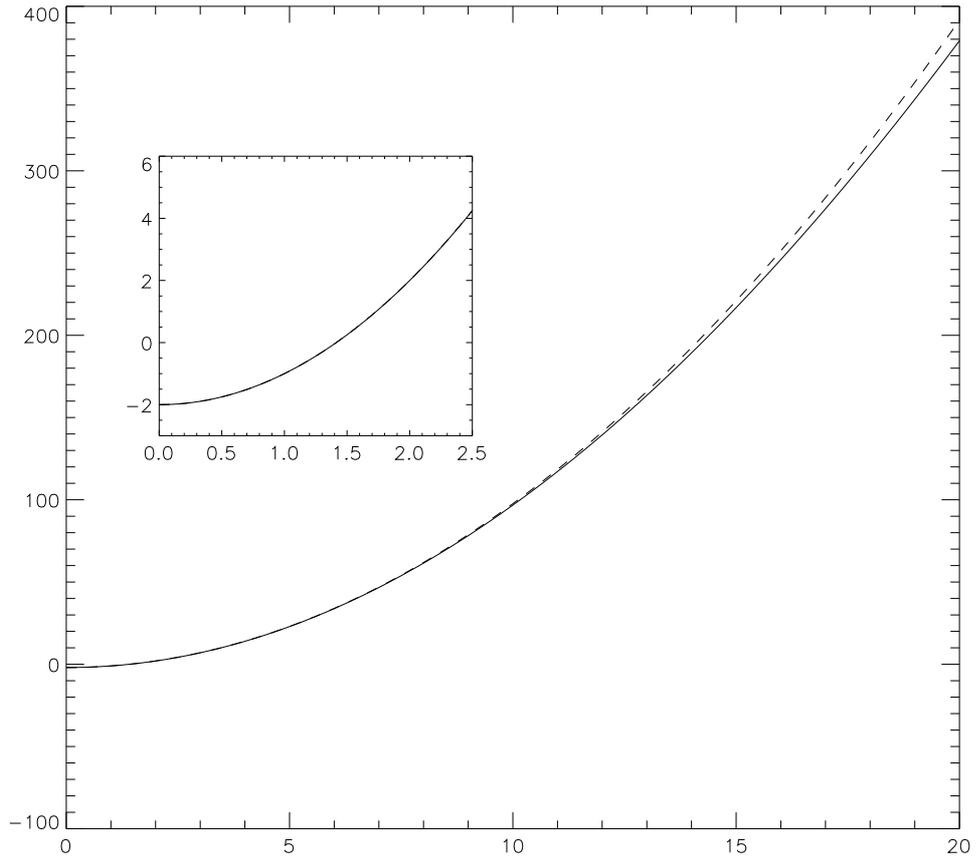

Figure 6: $\mathcal{O}_2(\phi)$ exact and perturbative after four renormalization group steps

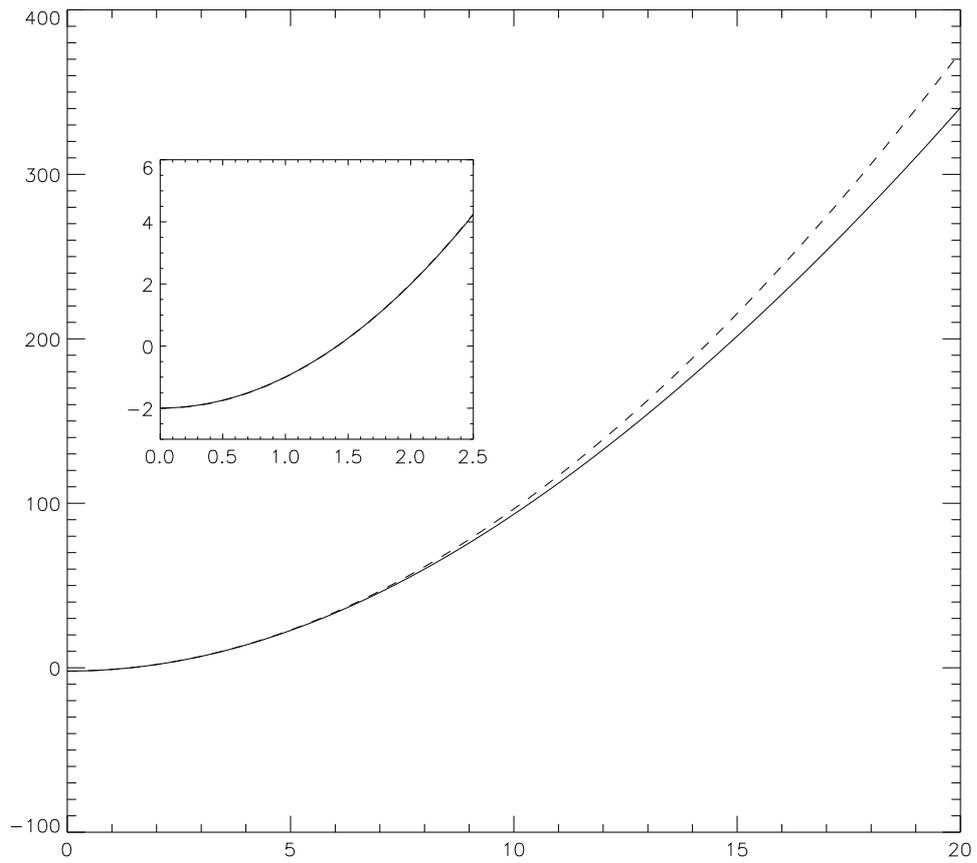

Figure 7: $\mathcal{O}_2(\phi)$ exact and perturbative after six renormalization group steps

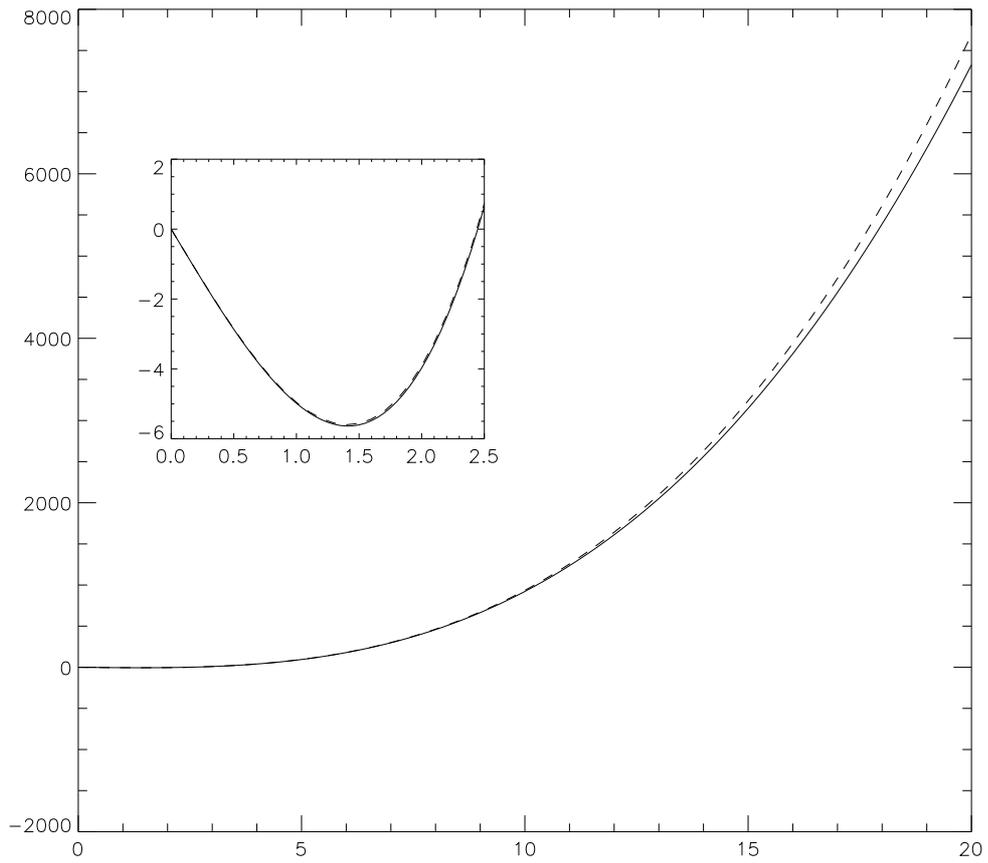

Figure 8: $\mathcal{O}_3(\phi)$ exact and perturbative after four renormalization group steps

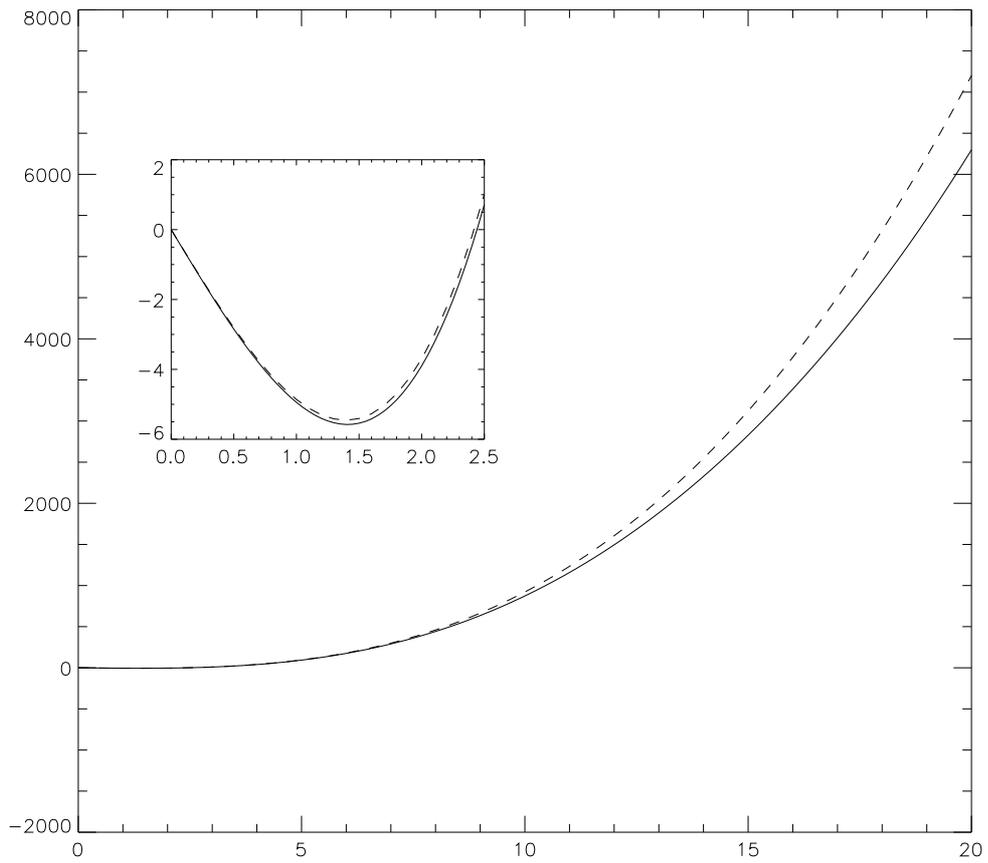

Figure 9: $\mathcal{O}_3(\phi)$ exact and perturbative after six renormalization group steps

33